\documentclass[pra,twocolumn,showpacs,superscriptaddress,groupedaddress,showpacs,shokeys]{revtex4}

\usepackage[utf8]{inputenc} 
\usepackage[T1]{fontenc}    
\usepackage{hyperref}       
\usepackage{url}            
\usepackage{amsmath}
\usepackage{amsfonts}
\usepackage{amssymb}
\usepackage{mathrsfs}
\usepackage{tikz}
\usepackage{subcaption}
\usepackage{graphicx}

\begin{document}

\title{Monte Carlo study of the BMV vacuum linear magnetic birefringence experiment}

\author{J. Agil\footnote{jonathan.agil@lncmi.cnrs.fr}}
\affiliation{Laboratoire
National des Champs Magn\'etiques Intenses (UPR 3228,
CNRS-UPS-UGA-INSA), F-31400 Toulouse Cedex, France}
\author{R. Battesti}
\affiliation{Laboratoire
National des Champs Magn\'etiques Intenses (UPR 3228,
CNRS-UPS-UGA-INSA), F-31400 Toulouse Cedex, France}
\author{C. Rizzo}
\affiliation{Laboratoire
National des Champs Magn\'etiques Intenses (UPR 3228,
CNRS-UPS-UGA-INSA), F-31400 Toulouse Cedex, France}

\begin{abstract}
QED vacuum can be polarized and magnetized by an external electromagnetic field, therefore acting as a birefringent medium.  This effect has not yet been measured.  In this paper, after having recalled the main facts concerning Vacuum Magnetic Birefringence polarimetry detection method and the related noise sources, we detail our Monte Carlo simulation of a pulsed magnetic field data run. Our Monte Carlo results are optimized to match BMV experiment 2014 data. We show that our Monte Carlo approach can reproduce experimental results giving an important insight to the systematic effects limiting experiment sensitivity.
\end{abstract}


\maketitle

\section{Introduction}
\subsection{Scientific Motivation}

In the framework of quantum electrodynamics (QED), a vacuum magnetic linear birefringence \cite{Battesti2013} is predicted.  This means that, in the presence of an applied magnetic field, $B_0$, the index of refraction for light polarized parallel, $n_{\parallel}$, and perpendicular, $n_{\perp}$, to the field can be written \cite{Battesti2013}

\begin{align}
n_{\parallel} &= 1+c_{0,2}\frac{B_0^2}{\mu_0}\label{eqn:n_parallel}\\
n_{\perp} &= 1+4 c_{2,0}\frac{B_0^2}{\mu_0}\label{eqn:n_perp}
\end{align}
with \cite{Battesti2013}
\begin{equation}\label{eqn:ccoef}
    c_{2,0} = \frac{2\alpha^{2}\hbar^{3}}{45m_\mathrm{e}^{4}c^{5}},\quad\quad  c_{0,2} = \frac{14\alpha^{2}\hbar^{3}}{45m_\mathrm{e}^{4}c^{5}},
\end{equation}
where $\mu_0$ is the vacuum permeability, $c$ is the speed of light in vacuum, $m_\mathrm{e}$ is the electron mass, $\alpha$ is the fine-structure constant and $\hbar$ is the reduced Planck's constant. These expressions come from the contribution of virtual electron-positron pair in vacuum and are valid for $B_0 \ll B_{cr}$ where $B_{cr}=m_e^{2}c^{2}/\hbar e$ is called the QED critical magnetic field.

This vacuum refractive index's dependence on $B_0^2$ is the signature of the $3^\mathrm{rd}$ order nonlinear Cotton-Mouton effect \cite{Shen}.  As first suggested in \cite{Erber}, one can therefore define a Cotton-Mouton constant for vacuum, $k_\mathrm{CM}$, that depends on the refractive index anisotropy, $\Delta{n}$, of vacuum
\begin{equation}
	\Delta{n} = n_{\parallel}-n_{\perp} = \frac{2\alpha^{2}\hbar^{3}}{15{\mu_0}m_\mathrm{e}^{4}c^{5}} B_0^2 = k_\mathrm{CM}B_0^2. \label{eqn:Delta_n_cm}
\end{equation}

The QED Cotton-Mouton constant of vacuum is $k_\mathrm{CM}\approx4\times10^{-24}$~T$^{-2}$. This effect remains yet unmeasured \cite{HIMAFUN}.

In 1979 Iacopini and Zavattini published a scheme to measure vacuum birefringence as an induced ellipticity, $\Psi$, in a laser field polarization as it passes through a vacuum where an external magnetic field is present~\cite{Iacopini1979}. Indeed, the ellipticity gained by the light going through the magnetic field region is given by 
\begin{equation}
\Psi=\pi\frac{L_B}{\lambda}\Delta{n}\sin 2\theta_P,
\label{eq:psiDn}
\end{equation} 
where $\lambda$ is the wavelength of the laser, $L_B$ the length of the magnetic field region and $\theta_P$ the angle between the light polarization and the magnetic field.

Currently, the existing vacuum birefringence polarimetry experiments are \textit{Bir\'{e}fringence Magn\'{e}tique du Vide} (BMV; Toulouse, France) \cite{Cadene2014} and \textit{Observing VAcuum  with  Laser} (OVAL; Tokyo, Japan) \cite{OVAL}. The BMV \cite{Cadene2014} and the OVAL \cite{OVAL} experiments are based on the use of pulsed magnets. Today, the best bound, $k_\mathrm{CM} =(1.9\pm 2.7) \times10^{-23}$~T$^{-2}$ with a coverage factor~\cite{GUM} $k=1$, has been published by the \textit{Polarizzazione del Vuoto con LASer} collaboration (PVLAS; Ferrara, Italy) \cite{DellaValle2016}, \cite{Ejlli2020}. This bound is about an order of magnitude higher than the QED predicted value (see also \cite{HIMAFUN}) and it has been obtained thanks to an acquisition time of a few $10^6$~s in a $B_0$ of 2.5~T provided by a permanent rotating magnet.

With improvements in optics, in particular in the field of very energetic laser sources, magnet technology, and interferometric technique, interest in the measurement of Vacuum Magnetic Birefringence (VMB) is ever-increasing \cite{HIMAFUN}. For example, recently, in~\cite{Karbstein2018} and~\cite{Shen2018} the observation of VMB of an X-ray beam interacting with a very powerful laser beam has been proposed. Efforts towards a direct measurement are also stimulated by the recent claim of an astronomical measurement of the degree of polarization of light from a neutron star resulting from vacuum birefringence~\cite{Astro-BMV}, \cite{Maiani2018}. Moreover, ATLAS and CMS collaborations have reported in 2019~\cite{Atlas2019}, \cite{CMS2019} a first observation of photon-photon scattering at very high energy, providing a new successful test of QED in the photon sector~\cite{Battesti2013}. From a more general point of view, let's note that VMB experiments contribute to particle physics in the realm of dark matter, axion and axion-like particles~\cite{HIMAFUN}, as well.

In this paper, after having recalled the main facts concerning VMB polarimetry experimental method and the related noise sources, we detail our Monte Carlo simulation of a pulsed magnetic field data run. Our Monte Carlo results are finally optimized to match our 2014 data \cite{Cadene2014}. We show that our Monte Carlo approach can reproduce experimental results giving an important insight to the systematic effects limiting experiment sensitivity.

\subsection{Experimental methods}

\begin{figure}
    \centering
\tikzset{every picture/.style={line width=0.75pt}} 
\resizebox{\linewidth}{!}{
\begin{tikzpicture}[x=0.75pt,y=0.75pt,yscale=-1,xscale=1]

\draw    (250,90) -- (630,90) ;
\draw    (370,60) -- (370,120) ;
\draw    (390,120) -- (370,120) ;
\draw    (390,60) -- (370,60) ;
\draw    (560,60) -- (560,120) ;
\draw    (538.09,119.96) -- (560,119.98) ;
\draw    (538.09,59.98) -- (560,60) ;
\draw    (390,60) .. controls (377.33,67.27) and (378,110.6) .. (390,120) ;
\draw    (538.09,59.98) .. controls (550.09,65.91) and (551.42,111.91) .. (538.09,119.98) ;
\draw   (570,79.98) -- (610,79.98) -- (610,99.98) -- (570,99.98) -- cycle ;
\draw    (570,99.98) -- (610,79.98) ;
\draw    (590,89.98) -- (610,49.98) ;
\draw   (595,69.98) .. controls (595,67.22) and (597.24,64.98) .. (600,64.98) .. controls (602.76,64.98) and (605,67.22) .. (605,69.98) .. controls (605,72.74) and (602.76,74.98) .. (600,74.98) .. controls (597.24,74.98) and (595,72.74) .. (595,69.98) -- cycle ;
\draw  [fill={rgb, 255:red, 0; green, 0; blue, 0 }  ,fill opacity=1 ] (601,69.98) .. controls (601,69.43) and (600.55,68.98) .. (600,68.98) .. controls (599.45,68.98) and (599,69.43) .. (599,69.98) .. controls (599,70.53) and (599.45,70.98) .. (600,70.98) .. controls (600.55,70.98) and (601,70.53) .. (601,69.98) -- cycle ;
\draw   (345,90) .. controls (345,87.24) and (347.24,85) .. (350,85) .. controls (352.76,85) and (355,87.24) .. (355,90) .. controls (355,92.76) and (352.76,95) .. (350,95) .. controls (347.24,95) and (345,92.76) .. (345,90) -- cycle ;
\draw  [fill={rgb, 255:red, 0; green, 0; blue, 0 }  ,fill opacity=1 ] (351,90) .. controls (351,89.45) and (350.55,89) .. (350,89) .. controls (349.45,89) and (349,89.45) .. (349,90) .. controls (349,90.55) and (349.45,91) .. (350,91) .. controls (350.55,91) and (351,90.55) .. (351,90) -- cycle ;
\draw   (290,80) -- (330,80) -- (330,100) -- (290,100) -- cycle ;
\draw    (620,90) -- (620,72) ;
\draw [shift={(620,70)}, rotate = 450] [color={rgb, 255:red, 0; green, 0; blue, 0 }  ][line width=0.75]    (10.93,-3.29) .. controls (6.95,-1.4) and (3.31,-0.3) .. (0,0) .. controls (3.31,0.3) and (6.95,1.4) .. (10.93,3.29)   ;
\draw  [fill={rgb, 255:red, 0; green, 0; blue, 0 }  ,fill opacity=1 ] (616.2,40.2) .. controls (616.2,39.65) and (615.75,39.2) .. (615.2,39.2) .. controls (614.65,39.2) and (614.2,39.65) .. (614.2,40.2) .. controls (614.2,40.75) and (614.65,41.2) .. (615.2,41.2) .. controls (615.75,41.2) and (616.2,40.75) .. (616.2,40.2) -- cycle ;
\draw  [fill={rgb, 255:red, 0; green, 0; blue, 0 }  ,fill opacity=1 ] (641.2,90) .. controls (641.2,89.45) and (640.75,89) .. (640.2,89) .. controls (639.65,89) and (639.2,89.45) .. (639.2,90) .. controls (639.2,90.55) and (639.65,91) .. (640.2,91) .. controls (640.75,91) and (641.2,90.55) .. (641.2,90) -- cycle ;
\draw   (624.92,45.18) .. controls (624.71,29.23) and (618.19,26.01) .. (605.39,35.53) ;
\draw   (640.06,101.01) .. controls (654.35,93.92) and (654.44,86.65) .. (640.32,79.23) ;
\draw    (465,130) -- (463.5,70) ;
\draw    (465,130) -- (466.5,70) ;
\draw    (470,70) -- (466.5,70) ;
\draw    (465,50) -- (460,70) ;
\draw    (465,50) -- (470,70) ;
\draw    (463.5,70) -- (460,70) ;

\draw    (178.67,73) -- (249.67,73) -- (249.67,108) -- (178.67,108) -- cycle  ;
\draw (176.67,79) node [anchor=north west][inner sep=0.75pt]  [font=\scriptsize]  {$ \begin{array}{l}
\text{Laser}\\
\lambda =1064\ \text{nm}
\end{array}$};
\draw (291.5,60.9) node [anchor=north west][inner sep=0.75pt]    {$P$};
\draw (571.5,60.9) node [anchor=north west][inner sep=0.75pt]    {$A$};
\draw (371.5,39.9) node [anchor=north west][inner sep=0.75pt]    {$M_{1}$};
\draw (540,39.9) node [anchor=north west][inner sep=0.75pt]    {$M_{2}$};
\draw (652.67,81.73) node [anchor=north west][inner sep=0.75pt]    {$Ph_{e}$};
\draw (610.67,9.73) node [anchor=north west][inner sep=0.75pt]    {$Ph_{t}$};
\draw (460,29.4) node [anchor=north west][inner sep=0.75pt]    {$\vec{B}$};

\end{tikzpicture}}
  \caption{Simplified diagram of the BMV experiment, where $P$ is the polarizer, $A$ the analyzer, $M_1$ and $M_2$ are the cavity mirrors and $\vec{B}$ is the magnetic field.}\label{fig:schemaexp}
\end{figure}
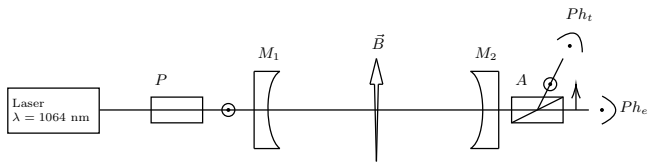

The principle of a polarimetry vacuum birefringence search is illustrated in Fig.~\ref{fig:schemaexp} following \cite{Battesti2008}. A laser field is linearly polarized by the polarizer prism $P$ at an angle of $\theta_P=45^{\circ}$ with respect to the direction of an applied transverse external field, $\vec{B}(t)$ to maximise the effect (see Eq.~\ref{eq:psiDn}). After the magnet, a second prism $A$ is used to analyze the polarization state of the outcoming light. The power of the ordinary beam $P_{t}$ and of the extraordinary one $P_{e}$ are monitored by the photodiodes $Ph_{t}$ and $Ph_{e}$. A Fabry-Perot high finesse cavity constituted by the mirrors $M_{1}$ and $M_{2}$ is used to increase the effect to be measured by trapping the light in the magnetic field region.

The Fabry-Perot cavity increases the effect due to a single pass in the magnetic field region by a factor $2\mathcal{F}/\pi$ where we have introduced the cavity finesse $\mathcal{F}$ \cite{born1980wolf}. This factor can be as high as 500 000 \cite{DellaValle2014} thanks to the progress in mirror fabrication.

The mirrors constituting the Fabry-Perot cavity show also an intrinsic birefringence that has been observed experimentally many times and whose origin is still not well understood \cite{MirrorBir}\cite{Xiao2020}. This causes an ellipticity, $\Gamma$, that has to be taken into account in the formula giving the ratio between the ordinary and extraordinary powers. $\Gamma$ can be related to the phase delay induced on the light by a single reflexion on the mirror reflecting layers by the following formula \cite{jacob1995supermirror}
\begin{equation}
    \Gamma=\frac{2\mathcal{F}}{\pi}\left(\frac{\delta_{1}}{2}\sin\left(2\left(\theta_{1}-\theta_{P}\right)\right)+\frac{\delta_{2}}{2}\sin\left(2\left(\theta_{2}-\theta_{P}\right)\right)\right),
    \label{eq:gamma}
\end{equation}
where $\delta_{i}$ are the phase delay induced by mirror $i$, and $\theta_{1}$, $\theta_{2}$ and $\theta_{P}$ the angles giving the orientation of the fast axis of first mirror, of the fast axis of the second mirror and of the light polarization with respect to a given reference system. The cavity birefringence $\Gamma$ can therefore be adjusted by varying $\theta_{1}$ and $\theta_{2}$ by rotating the mirrors (see \textit{e.g.} \cite{Cadene2014}).

Taking into account also that polarizers are not ideal and that they show an extinction ratio $\sigma^{2}$ (see {\it e.g.}~\cite{Takudo1998}), the power ratio is given by
\begin{align}
    \frac{P_{e}(t)}{P_{t}(t)}&=\sigma^{2}+\left[\Gamma+\Psi(t)\right]^{2}\\
    &\simeq\sigma^{2}+\Gamma^{2}+2\Gamma\Psi(t),\label{eq:Psi2014}
\end{align}
when $\Psi(t)\ll\Gamma$,  $\sigma^2$ and $\Gamma$ are assumed constant in time and any noise contribution is neglected.

A crucial issue for the data analysis is the fact that a Fabry-Perot cavity shows a different dynamical behavior in the ordinary outgoing beam than in extraordinary one. $P_t$ time response is typical of a first order low pass filter,
\begin{equation}
    H(\nu)=\frac{1}{1+i\nu/\nu_{c}},
\end{equation}
with $\nu_{c}=c/4L_{c}\mathcal{F}$ with $L_{c}$ the cavity length. On the other hand, the extraordinary beam reacts differently, as if the cavity had filtered it twice.  The cavity transfer function seen by the extraordinary beam is \cite{berceau2010dynamical}
\begin{equation}
    H^{\prime}(\nu)=\frac{1}{\left(1+i\nu/\nu_{c}\right)^{2}},
\end{equation}

In order to work out the formula giving $P_e$ in function of the optical parameters, we introduce the operator $\hat{C}$ which is associated to the transfer function of the cavity so that $\hat{C}[f(t)]=\mathscr{F}^{-1}[H(\nu)\mathscr{F}[f(t)]]$ where $\mathscr{F}$ is the operator associated with the Fourier transform.

$P_{e}(t)$ has to be written as
\begin{equation}
    P_{e}(t) =\sigma^{2}\hat{C}\left[\eta P_i(t)\right] + \hat{C}\left[\hat{C}\left[\eta P_i(t)\right]\left(\Gamma+\Psi(t)\right)^{2}\right],
    \label{eq:psi}
\end{equation}
where $P_i(t)$ is the light power injected into the cavity, $\eta$ corresponds to the coupling of the cavity and $\sigma^{2}$ is assumed to be constant. Since $P_t(t)=\hat{C}\left[\eta P_i(t)\right]$
\begin{equation}
   \Psi(t)=\sqrt{\frac{ \hat{C}^{-1}\left[P_{e}(t) - \sigma^{2}P_t(t)\right]}{P_t(t)}} - \Gamma.
    \label{eq:psi1}
\end{equation}

Assuming that the time variation of $P_t(t)$ can be neglected as well, formula \ref{eq:psi} can be simplified as
\begin{equation}
    P_{e}(t) =\sigma^{2}P_t + P_t \hat{C}\left[\left(\Gamma+\Psi(t)\right)^{2}\right],
    \label{eq:psi2}
\end{equation}
which, when $\Psi(t) \ll \Gamma$, gives
\begin{equation}
   \frac{ \frac{P_{e}(t)}{P_t}  - \sigma^{2} - \Gamma^2}{2\Gamma} \approx \hat{C}\left[\Psi(t)\right] \propto \hat{C}\left[B(t)^2\right].
    \label{eq:psi3}
\end{equation}

Let's note that neither Eq. \ref{eq:psi1} nor Eq. \ref{eq:psi3} has been used in ref. \cite{Cadene2014}. We discuss in the appendix the implication of that.

Typically, the pulse of the magnetic field can be written as an oscillation strongly dumped
\begin{equation}
    B(t)\propto\sin(2\pi \nu t)e^{-t/\tau}.
\end{equation}

\begin{figure}
    \centering
    \includegraphics[width=\linewidth]{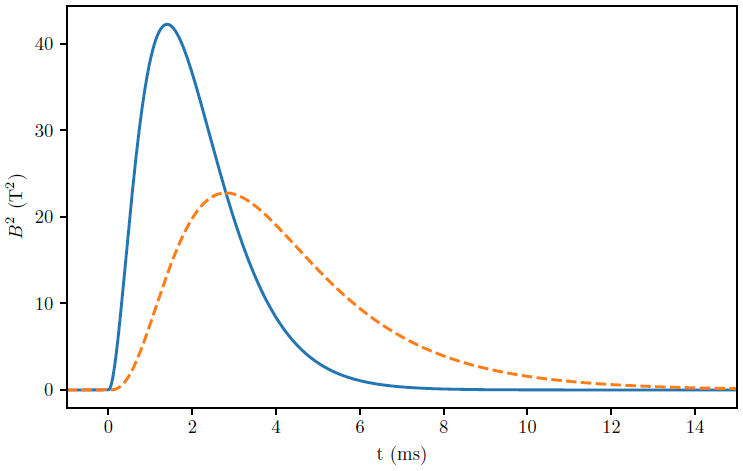}
    \caption{In full blue line, the square of the modeled magnetic field $B^{2}(t)$ and in dashed orange line the $B^{2}(t)$ field filtered by a first order low pass filter of 74~Hz bandwidth.}
    \label{fig:B2}
\end{figure}

In Fig. \ref{fig:B2} we show a $B^{2}(t)$ pulse that simulates the one of ref. \cite{Cadene2014}. The maximum of the field is $B_{max}=6.5$~T, $\nu=5$~Hz et $\tau=1.4$~ms and to calculate the filtered $B^{2}(t)$ we have assumed that the cavity bandwidth is $\nu_{c}=74$~Hz, again like in ref. \cite{Cadene2014}. Following Eq.~\ref{eq:psi3} the signal to look for by monitoring $P_e/P_t$ is proportional to the filtered $B^{2}(t)$ showed in Fig.~\ref{fig:B2}. The magnet pulse is obtained by charging a bank of capacitor \cite{HIMAFUN}. For a pulse as  the one of Fig.~2 the bank of capacitor is charged with an energy of a few 100~kJ of energy. The pulse also generate, by the energy discharge, sound propagating in air or in the apparatus.

\subsection{Noise evaluation}

For each sample, the signal to be measured $S_s$ is equal to $2P_t\Gamma \Psi(t) T_{i}q/h\nu$ where $\nu$ is the laser frequency, $T_i$ is the sampling period and $q$ is the photodiode quantum efficiency. On the other hand, the shot noise is calculated from the DC component of $P_e$, $P_{DC}= P_{t}\left(\Gamma^2+\sigma^2 \right )$, and therefore $S_n = \sqrt{P_t \left(\Gamma^2+\sigma^2  \right ) T_{i}q/h\nu}$.

We can thus infer, in the ideal case of a shot noise limited experiment, the signal to noise ratio is
\begin{equation}\label{StoNpol}
\frac{S_s}{S_n} = \sqrt{\frac{4P_t\Gamma^2\Psi^2(t)}{h\nu\left(\Gamma^2 + \sigma^2\right)} T_{i}q},
\end{equation}
and in the case $\Gamma \gg \sigma$ the sensitivity $\Psi_s$ \textit{i.e.} the ellipticity measurable with one second of total integration time $T_t=N T_i$ where $N$ is the number of pulses, with a signal to noise ratio of 1 is
\begin{equation}\label{Sensbir}
    \Psi_s = \sqrt{\frac{h\nu}{4qP_t}},
\end{equation}
as in ref. \cite{Battesti2008}.

In terms of variation of the refractive index, at the maximum field, taking into account that a Fabry-Perot cavity acts as a low pass filter \cite{Battesti2008},\cite{Berceau2012}, the signal to noise ratio can be written as
\begin{equation}\label{StoNnuM}
    \frac{S_s}{S_n} =  \frac{1}{\sqrt{\nu_c^2+\nu_e^2}} f_f k_{\mathrm{CM}} B_{max}^{2}\sqrt{\frac{q P_t \nu T_i}{h}},
\end{equation}
where the filling factor $f_f=L_{B}/L_{c}$, $\nu_e$ is the effect modulation frequency, $\nu_c = c/4 L_c \mathcal{F}$ is the cavity filter bandwidth and $B_{max}$ the maximum of the magnetic field reached during a pulse.
In the case of a pulsed magnetic field the definition of $\nu_e$ is not straightforward because the $B(t)$ is not an harmonic function. To apply Eq.~\ref{StoNnuM} to pulsed case, let's assume that for pulsed magnets and for a given cavity bandwidth $\nu_c$, $\nu_e$ is defined by the equation
\begin{equation}
\max[\hat{C}(B^{2}(t))]=\frac{1}{\sqrt{1+\frac{\nu^{2}_e}{\nu^{2}_c}}}B^{2}_{max}.
\end{equation}
In the case of Fig.~\ref{fig:B2} where $\nu_c=74$~Hz, following this definition, $\nu_e$ is equal to 115 Hz.

In the best case, the total integration time needed to reach a signal to noise ratio equal to 1 can be written as
\begin{equation}\label{Tsn}
    T_t\left(\frac{S_s}{S_n}=1\right) = \frac{\nu_c^2+\nu_e^2}{f_f^2 k_{\mathrm{CM}}^2B_{max}^4}\frac{h}{\nu q P_t}.
\end{equation}

To estimate $T_t\left(\frac{S_s}{S_n}=1\right)$ let us assume some typical values for the experimental parameters : $\nu_c = \nu_e = 100$~Hz, $f_f = 0.1$, $k_{\mathrm{CM}} = 4\times 10^{-24}$~T$^{-2}$, $B_{max}=10$~T, $q=0.5$, $P_t = 1$~W. The resulting  $T_t\left(\frac{S_s}{S_n}=1\right)$ is about 50~s. But experiments as BMV or PVLAS are not shot noise limited. The reasons given to explain this fact are related to mirror birefringence noise caused by mirror position noise \cite{Hartman2017}, mirror temperature induced noise \cite{PVLASNoise} and/or noise sources proportional to the cavity finesse \cite{Ejlli2020}.

In the general case, the expression for the ellipticity noise $\delta\Psi$ is obtained by $\delta P$, the noise associated to the $P$ power where $P$ is the most general expression for the power detected by the photodiode at the extinction. This noise is constituted by noise associated to the experimental parameters on which $P_e$ depends plus $P_e$ shot noise, $\delta P_{shot}$, that we have calculated before plus  $\delta P_{o}$ which represents any noise whose cause is not explicitly in the $P_e$ formula. Let's note that for any variable $v$, its noise is indicated as $\delta v$.

From the expression of the DC component of $P_e$ we have
\begin{align}
\delta P_{DC}&=\delta P_t(\sigma^{2}+\Gamma^{2})+P_t(\delta \sigma^{2}+\delta\Gamma^{2})\nonumber\\
&=\delta P_t(\sigma^{2}+\Gamma^{2})+P_t(\delta \sigma^{2}+2\Gamma\delta\Gamma),
\end{align}
thus we can derive $\delta\Psi=\delta P/2P_t\Gamma$ following the  expressions
\begin{align}\label{DeltaPsiPolHo}
\delta\Psi =& \frac{\delta P_{DC}+\delta P_{shot}+\delta P_o}{2P_t\Gamma}\nonumber\\
 \nonumber  ={}& \frac{\delta P_t}{2P_t} \left( \frac{\sigma^2}{\Gamma} + \Gamma \right) + \frac{\delta\sigma^2}{2\Gamma} \\
 &+ \delta\Gamma + \frac{\delta P_{shot}}{2P_t\Gamma} + \frac{\delta P_o}{2P_t\Gamma}.
\end{align}
The $\delta P_t$ noise, corresponding here to the power noise after the Fabry-Perot cavity, appears as an ellipticity noise since it beats with all the polarization parameters. Obviously, noise in these optical parameters becomes also of a crucial importance.

\section{Monte Carlo study}

Our Monte Carlo simulation aims to reproduce real data as the one of ref.~\cite{Cadene2014} to evaluate what are the sources of noise affecting mostly the BMV experiment. Secondly, once we are able to simulate a set of data which mimic the real ones, we want to check our analysis procedures like the 2014 one.

\begin{figure}
    \centering
\includegraphics[width=\linewidth]{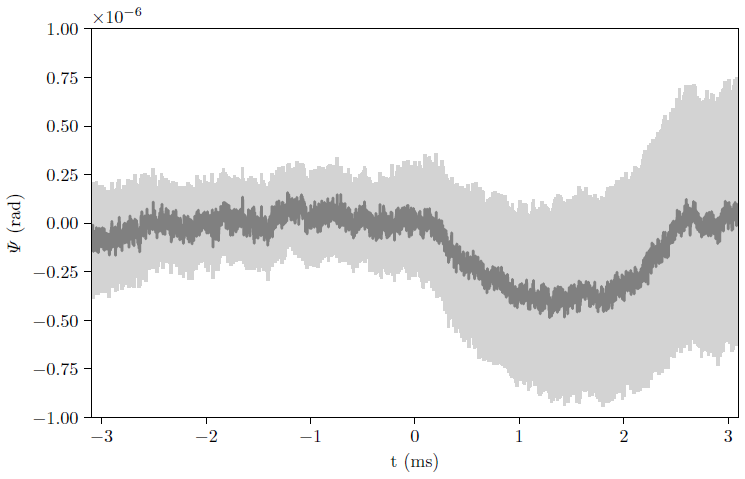}  
\caption{Time evolution of the ellipticity and its uncertainties with a coverage factor $k=3$.}
    \label{fig:don2014}
\end{figure}

In Fig. \ref{fig:don2014}, taken from ref.~\cite{Cadene2014}, we show the data that we have to reproduce first by our Monte Carlo method. This graph shows ellipticity measured by averaging around 100 pulses. At every instant, from -3.1~ms to +3.1~ms, a mean value and its uncertainty are associated at any time $t$. The instant $t=0$ corresponds to the beginning of the magnet pulse. Due to the pulse, there are two time domains $t<0$ and $t>0$ which have to be treated differently.

Moreover, let's recall that pulses in ref.~\cite{Cadene2014} were labeled following the sign of $\Gamma$ and of $B(t)$. The averaging to obtain the final data was done with the following formula
\begin{equation}
    \Psi(t)=\frac{1}{4}\Psi_{+,+}(t)+\frac{1}{4}\Psi_{+,-}(t)-\frac{1}{4}\Psi_{-,+}(t)-\frac{1}{4}\Psi_{-,-}(t),
    \label{eq:moy}
\end{equation}
where $\Psi_{\pm,\pm}$ represent the mean of four set of data with different sign of $\Gamma$ and of $B(t)$, the first subscript refering to the sign of $\Gamma$, the second of $B(t)$. In this equation, we use the fact that $\Psi$ is odd with respect to the sign of $\Gamma$ and even to the sign of $B(t)$. Thus to simulate it one has to simulate the four type of raw data $\Psi_{\pm,\pm}$. In the following, we therefore generate 25 random signals for any class of $\Psi_{\pm,\pm}$ and take the mean to get the four signals. The Monte Carlo final data is obtained by the Eq.~\ref{eq:moy}.

Another important feature of the 2014 data \cite{Cadene2014} is that almost half of the pulses has not been used for the final analysis because their distribution of $\Psi(t)$ values for $t<0$ were not gaussian. This was considered by the authors of ref.~\cite{Cadene2014} as an indication of some extra noise affecting the data even before the magnetic pulse. 

Our Monte Carlo data depend on some random variable whose interval of variation has to be optimized by comparing it with the target experimental data discussed before. The ideal goal is that, at any instant $t_i$ Monte Carlo average value $\Psi(t_i)_{\mathrm{MC}}$ and its uncertainty $\sigma(t_i)_{\mathrm{MC}}$ equalize the corresponding  $\Psi(t_i)_{exp}$ and $\sigma(t_i)_{exp}$ given by the target experiment. To optimize the Monte Carlo parameters we minimize the following quantity $O$
\begin{widetext}
\begin{equation}\label{Oeq}
    O = \sqrt{\frac{\sum_{i}\left[\left(\Psi(t_i)_{\mathrm{MC}} + \sigma(t_i)_{\mathrm{MC}}) - (\Psi(t_i)_{exp} + \sigma(t_i)_{exp}\right)\right]^2 + \sum_{i}\left[\left(\Psi(t_i)_{\mathrm{MC}} - \sigma(t_i)_{\mathrm{MC}}) - (\Psi(t_i)_{exp} - \sigma(t_i)_{exp}\right)\right]^2}{2}}.
\end{equation}
\end{widetext}
The way we write our optimization parameter $O$ corresponds to perform two least square fit at a time, one on the up values of the experimental data with their uncertainties and one on the bottom ones. Developing Eq.~\ref{Oeq} one obtains a more simple expression

\begin{equation*}\label{OeqSim}
    O = \sqrt{\sum_{i}\left[\left(\Psi(t_i)_{\mathrm{MC}} - \Psi(t_i)_{exp}\right)^2 + \left(\sigma(t_i)_{\mathrm{MC}} - \sigma(t_i)_{exp}\right)^2\right]}
\end{equation*}

which is equivalent to fit separately at any instant $t_i$ the average value $\Psi(t_i)_{exp}$ and the uncertainty $\sigma(t_i)_{exp}$.

\begin{figure}
    \centering
	\includegraphics[width=\linewidth]{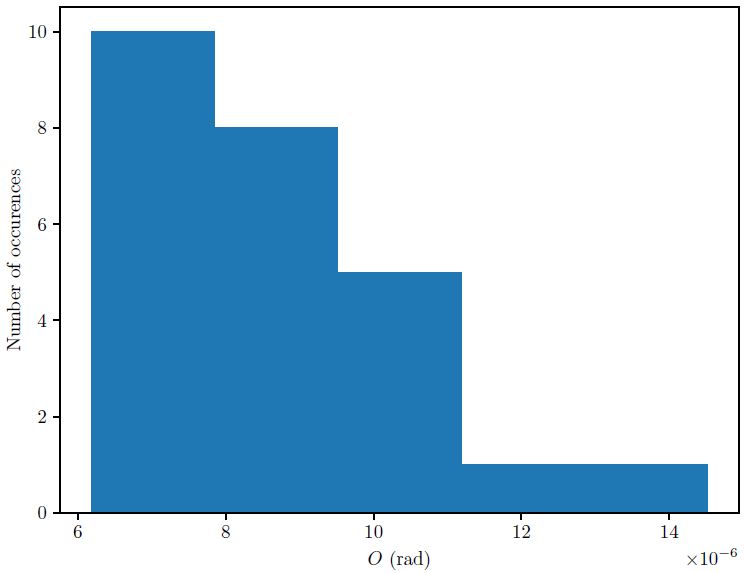}    
 \caption{Histogram of the $O$ values for the point $t_a=0.14$~ms of Fig.~\ref{fig:optipara}a).}
    \label{fig:histO}
\end{figure}

Our optimization procedure has also to take into account that for any fixed set of parameters, like the ranges of random variables, the $O$ parameter also shows a random distribution. In the following, for any set of parameters we simulate 25 experiments and we give the minimum value $O_{min}$. As we are near this minimum, the distribution of the $O$ values cannot be gaussian as shown in Fig.~\ref{fig:histO}. To characterize the extent of the $O$ values we define a quantity $\Delta O$, the interval of variation of $O$, obtained as
\begin{equation}\label{eq:DeltaO}
    \Delta O = \sqrt{\frac{1}{N-1}\sum_{l=0}^{N-1}\left( O_l - O_{min}\right)^2},
\end{equation}
where $N$ is number of $O_l$ value.

The optimized value of a free parameter is the one that gives the smallest $O_{min}$ with a $\Delta O$ associated that is also as small as possible.

\subsection{Data simulation for $t<0$}

\begin{figure}
    \centering
	\includegraphics[width=\linewidth]{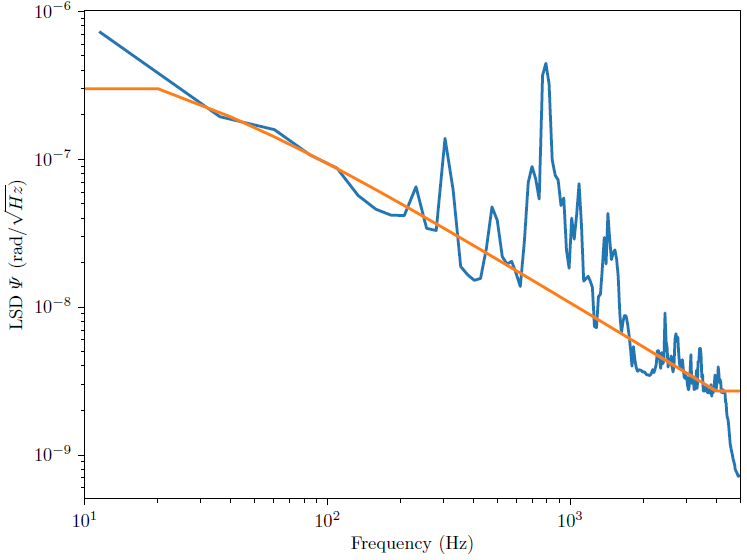}   
 \caption{Linearized Power Spectral Density of the ellipticity without a magnetic shot as given in \cite{Cadene2014} averaged to have a spectral resolution of about 20~Hz and its modeling.}
    \label{fig:Bruit}
\end{figure}

To reproduce data unperturbed by the magnetic pulse, we have to inject in our Monte Carlo simulation an expression for the $\delta \Psi$ noise. In Fig.~\ref{fig:Bruit} we show the linearized power spectral density (LSD) $\beta(\nu)$ that we have used and that is represented by the function
\begin{equation}\label{eq:LSDmodel}
    \beta(\nu)=\begin{cases}
       A a/(a+\nu) & \text{if $\nu\leq\nu_{0}$}\\
       A a/(a+\nu_{0}) & \text{if $\nu>\nu_{0}$}\\
    \end{cases},
\end{equation}
where
\begin{align}
    A&=\beta(\nu_{1})\frac{\nu_{0}-\nu_{1}}{\nu_{0}-\nu_{1}\frac{\beta(\nu_{1})}{\beta(\nu_{0})}},\\
    a&=\nu_{0}\frac{\beta(\nu_{0})}{A-\beta(\nu_{0})}\nonumber,
\end{align}
and $\nu_{1}=20$~Hz, $\nu_{0}=4000$~Hz, $\beta(\nu_{1})=3\times 10^{-7}$~rad/Hz$^{1/2}$ et $\beta(\nu_{0})=2.7\times 10^{-9}$~rad/Hz$^{1/2}$. The corresponding LSD is compared in Fig.~\ref{fig:Bruit} to the experimental LSD reported in ref.~\cite{Cadene2014}, to show mechanical resonances of the 2014 apparatus, averaged to have the same spectral resolution of about 20~Hz.

According to \cite{Timmer1995}, to generate a signal $N(t)$ in the time domain using a LSD we need to randomize the phase as well as the amplitude of each frequency, this is done as follows
\begin{equation}
    N(t)=\sum_{n=1}^{n_{max}}r\times \sqrt{df} \beta(\nu_{n})\cos\left( 2\pi\nu_{n} t+\varphi_{n}\right),
\label{eq:bruitmodif}
\end{equation}
where the phase $\varphi_{n}$ is a random variable taken uniformly in the interval $[-\pi,\pi[$, $r$ is also a uniformly distributed random variable in the interval $[1,1+\Delta r[$ and $n_{max}$ determine the maximum frequency $\nu_{max}=200$~kHz of the sum that is chosen to save computing time.
We have to introduce the $r$ random variable because between each pulses, experimentalists adjust optics obtaining every time a value of $\Gamma$ which is slightly different (see Eq.~\ref{eq:gamma}). There is no reason that this variable has to follow a Gaussian distribution so we assume that it is uniformly distributed. This is the main difference between this work and this of \cite{Timmer1995} where the amplitude follow a Gaussian distributions. As shown in Eq.~\ref{DeltaPsiPolHo} noise changes as a function of $\Gamma$. When $\delta P_t$ noise dominates, the overall noise increases when $\Gamma$ increases. 

\begin{figure}
\centering
\includegraphics[width=\linewidth]{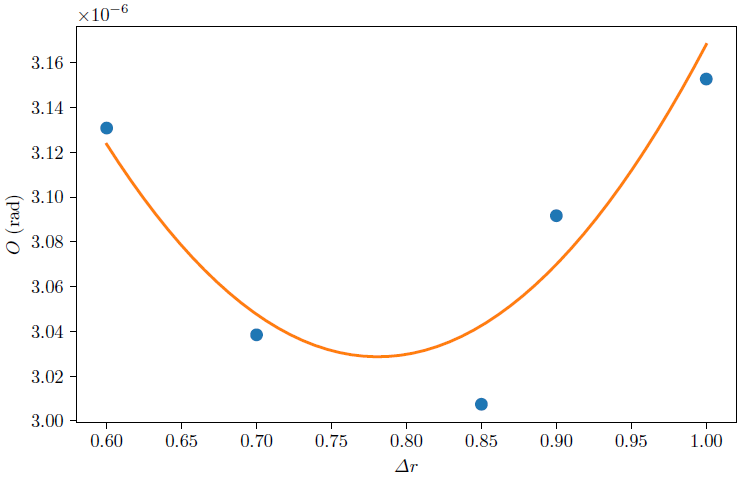}
\caption{Optimization parameter $O$ as a function of the $\Delta r$ parameter. A fit by a quadratic function gives a minimum of $\Delta r_{min}=0.78$.}
\label{fig:optiR}
\end{figure}

The value of $\Delta r$ has been optimized by calculating for a few value of $\Delta r$ and $O_{min}$ and the corresponding $\Delta O$ calculated with Eq.~\ref{eq:DeltaO}. We then perform a least square fit with a quadratic function to determine the position of the minimum. For each point we use $\Delta O$ as the weight for the fit. We obtain the value $\Delta r=0.78$. In Fig.~\ref{fig:optiR} we show the values of $O$ as a function of $\Delta r$ obtained thanks to our Monte Carlo simulation.

\begin{figure}
    \centering
    \begin{subfigure}{\linewidth}
    \centering
	\includegraphics[width=\linewidth]{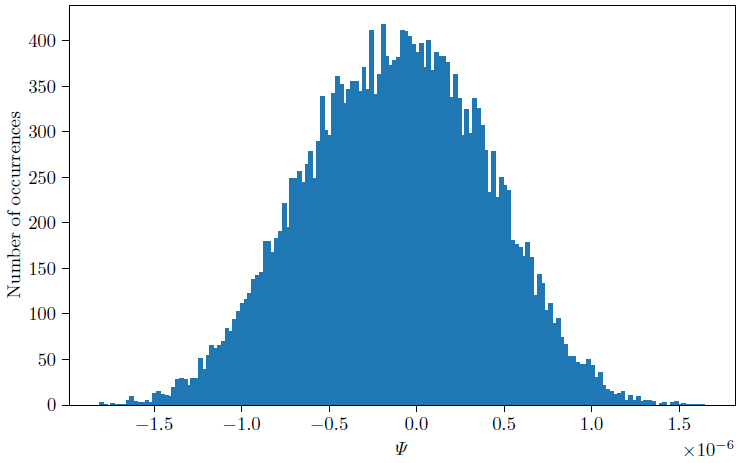}   
    \caption{}
    \label{fig:histpsia}
\end{subfigure}

\begin{subfigure}{\linewidth}
    \centering
    \includegraphics[width=\linewidth]{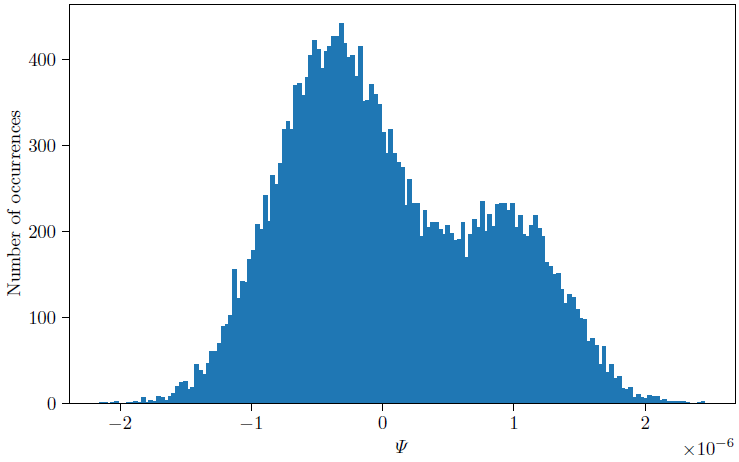}    
    \caption{}
    \label{fig:histpsib}
    \end{subfigure}
    \caption{Typical histogram of $\Psi(t)$ before the magnetic pulse. (a) Shot respecting a gaussian distribution. (b) Shot not showing a gaussian distribution. These results are to be compared with Fig.~5 of \cite{Cadene2014}.}
    \label{fig:histpsi}
\end{figure}

Our simulation can also be used to test $\Psi(t)$ distribution for $t<0$ of each simulated pulse. In Fig. \ref{fig:histpsia} and \ref{fig:histpsib} we show some Monte Carlo results simulating those presented in Fig.~5 of \cite{Cadene2014} indicating that we are able to reproduce the feature of experimental 2014 data without adding any other noise mechanism than a LSD showing that there were no extra noise before the shot. In principle, the rejected experimental data could have been taken into account in \cite{Cadene2014}. Thus in our simulation we will not put aside data not presenting a Gaussian distribution for $t<0$ as it was done in 2014.

\subsection{Data simulation for $t>0$}

For $t>0$, 2014 data set Fig.~\ref{fig:don2014} shows a complex structure that we can summarize as a sinusoidal time variation with an associated ever increasing error bar. The Monte Carlo random function becomes more complex as well. After testing several formulas for $N(t>0)$ \textit{i.e.} several different noise contributions, whose Monte Carlo result was so far from target data that they were not worth an optimization of their free parameters, we have hold the following expression as the more appropriate one
\begin{align}
    N(t)={}&\sum_{n=1}^{n_{max}}\left[r+\rho(t-t_a)\right]\sqrt{df} \beta(\nu_{n})\cos\left( 2\pi\nu_{n} t+\varphi_{n}\right)\nonumber \\ 
    &- A\sin\left(2\pi\nu_{a} (t-t_a)\right),
\label{eq:bruitmodifaugmenteavect}
\end{align}
where the amplitude of each frequency increases linearly from the value before the magnetic shot, starting at $t=t_a$ with slope $\rho$ which is a random variable in the interval $[0, \rho_{max}]$. We also add a sinusoidal oscillation of frequency $\nu_a$ and of random amplitude $A$ in the interval $[A_{acous}(1-r_{acous}),A_{acous}(1+r_{acous})]$ also starting at $t=t_a$. For $t<t_a$, we have that $A=0$ and $\rho=0$.
The physical meaning of Eq.~\ref{eq:bruitmodifaugmenteavect} is that the  acoustic perturbation generated by the pulse discharge arrives at a time $t_a$ and it can be represented by a periodic function of frequency $\nu_a$. It induces an increase of the noise level through the LSD that we approximate with a linear dependence on time.

\begin{figure*}
\centering
\includegraphics[width=\linewidth]{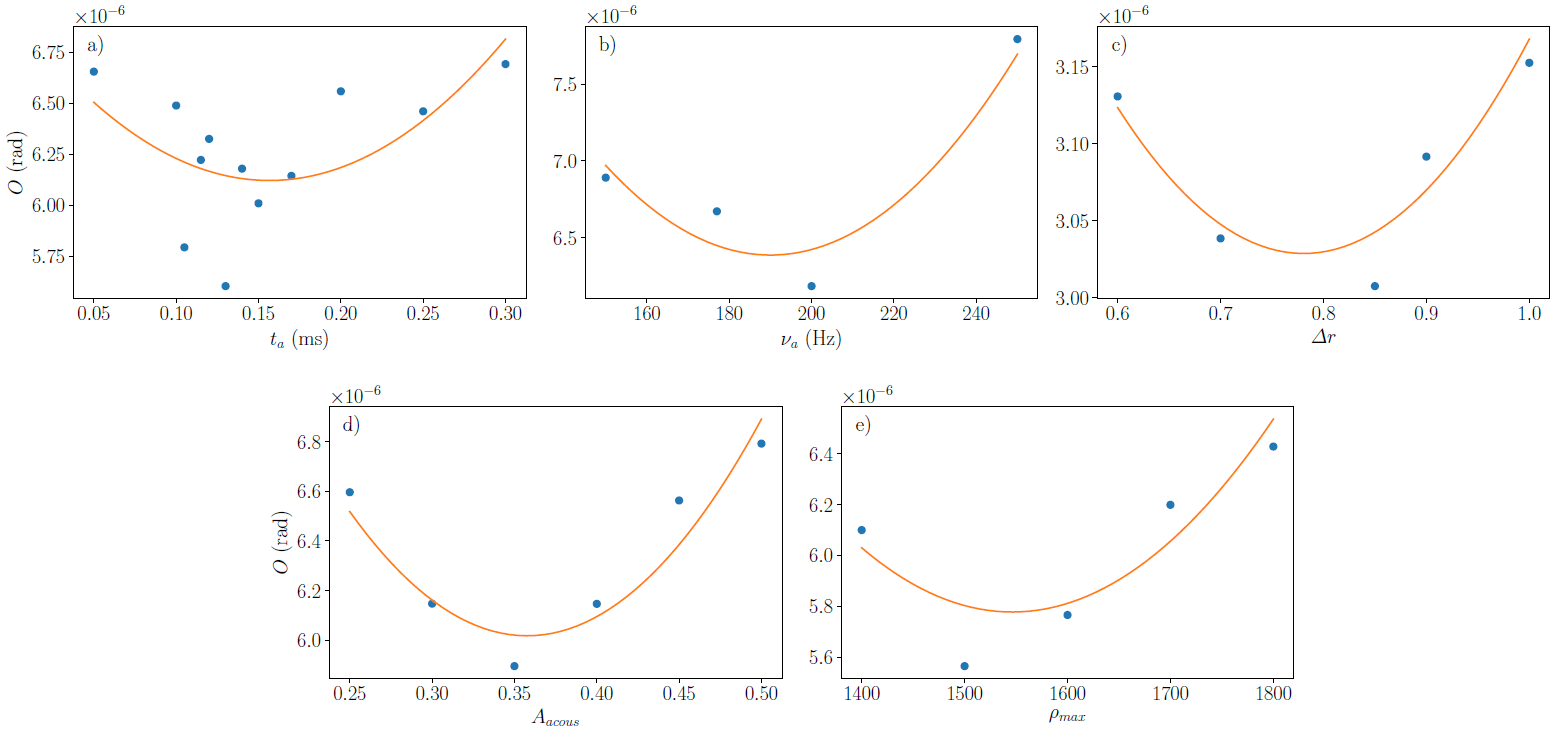}
\caption{Optimization parameter $O$ as a function of the Monte Carlo parameters, a least square fit by a quadratic function is performed to find the optimum. a) $t_{a_{min}}=0.157$~ms b) $\nu_{a_{min}}=190$~Hz c) $r_{acous_{min}}=3.22$ d) $A_{acous_{min}}=-0.36\times 10^{-6}$ e) $\rho_{max_{min}}=1550$.}
\label{fig:optipara}
\end{figure*}

We proceed with the optimization of the free parameters $\rho_{max}$, $A_{acous}$, $r_{acous}$, $\nu_a$ and $t_a$.
We start from some initial guesses for the parameters and we optimize them starting from the most impacting effect: first $t_a$ next $\nu_a$ then $r_{acous}$, $A_{acous}$ and finally $\rho_{max}$. For each parameter, we calculate a few of $O$ data point and perform a fit by a quadratic function to determine a minimum. For each data point we use its $\Delta O$ as a weight for the fit. The results are regrouped in Fig.~\ref{fig:optipara}.

\begin{figure}
\centering
\includegraphics[width=\linewidth]{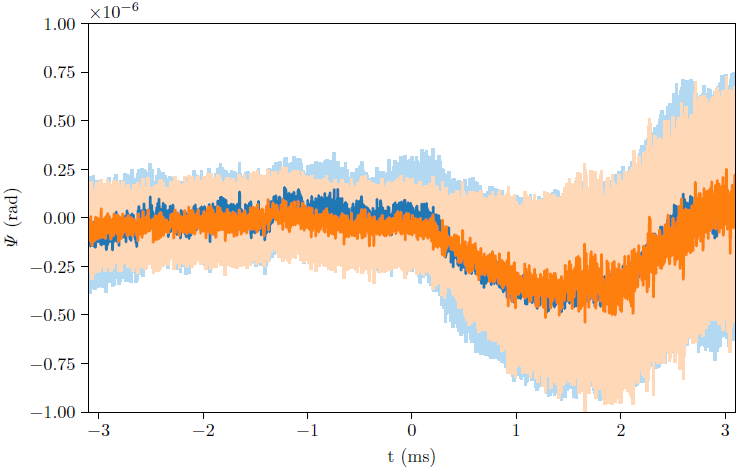}
\caption{Monte Carlo ellipticity in function of time. In blue, experimental data from \cite{Cadene2014} and its uncertainties with a coverage factor $k=3$. In orange, simulation presenting the minimal $O$ parameter with optimized Monte Carlo parameters and its uncertainties with a coverage factor $k=3$.}
\label{fig:optiresult}
\end{figure}

Once we optimized all of our parameters, we performed a second iteration of 25 computation with the optimized parameters and we saved the one with the lowest value of $O$. We can then compare the ellipticity obtained with the experimental data as shown in Fig.~\ref{fig:optiresult}. 

\begin{figure}
\centering
\includegraphics[width=\linewidth]{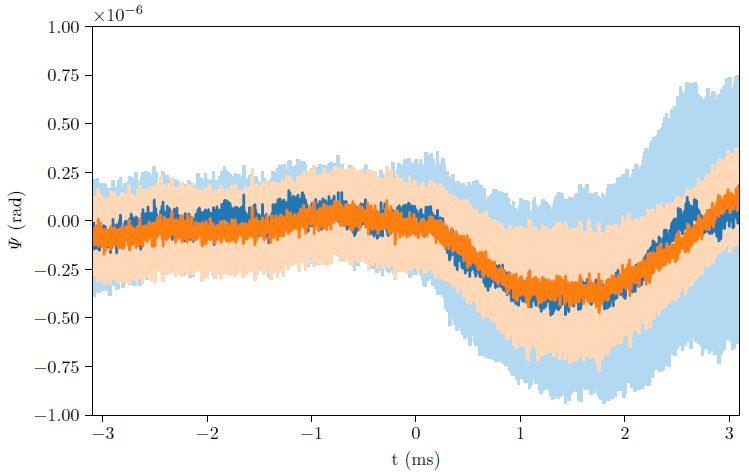}
\caption{Ellipticity in function of time. In blue, experimental data from \cite{Cadene2014} and its uncertainties with a coverage factor $k=3$. In orange, example of simulation without a random linear increase of the noise level and its uncertainties with a coverage factor $k=3$.}
\label{fig:sanslin}
\end{figure}

To show that systematic effects that we have introduced are necessary we show in Fig.~\ref{fig:sanslin} a result without the random linear increase of noise with time, that is to say $\rho_{max}=0$. We clearly observe that the uncertainties doesn't match with the experimental one.

\subsection{Analysis}
\begin{figure}
\centering
\includegraphics[width=\linewidth]{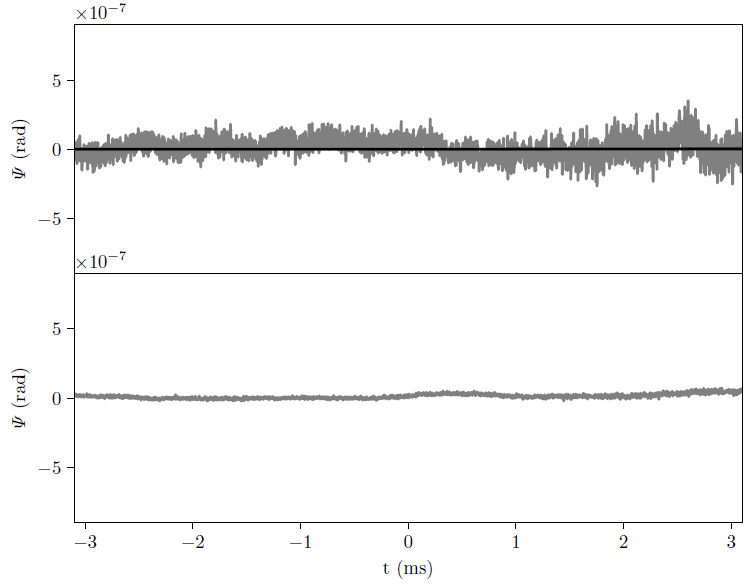}
\caption{Above: ellipticity obtained by substracting from the experimental data our Monte Carlo result with superimposed a least square fit of the Cotton-Mouton effect giveing the value $k_{\mathrm{CM}}=(-1.2\pm2.1)\times 10^{-21}$~T$^{-2}$ with a coverage factor $k=3$. Below: uncertainties obtained by substracting from the experimental uncertainties our Monte Carlo ones.}
\label{fig:residus}
\end{figure}

Once we have a set of optimized parameters and a Monte Carlo method mimicking real data we can use this machinery to improve data analysis. Since nothing in our simulation depends on $B^2(t)$ we are sure that the signal isn't being simulated as well. Actually we have simulated only noise sources affecting the experimental data, we therefore substract from the experimental data the result of our simulation and then estimate the Cotton-Mouton constant of vacuum by performing a fit with $B^2_f(t)$. We obtain a Cotton-Mouton constant $k_{\mathrm{CM}} = (-1.2\pm2.1)\times 10^{-21}$~T$^{-2}$ with a coverage factor $k=3$ (see Fig.~\ref{fig:residus}). We recall that in ref.~\cite{Cadene2014} the reported Cotton-Mouton constant obtained through a fit of the residuals of the fit of the systematic effect was $k_{\mathrm{CM}} = (5.1\pm6.2)\times 10^{-21}$~T$^{-2}$ with a coverage factor $k=3$. We observe that using our Monte Carlo we obtain a value that is more accurate since the value is compatible with zero with a coverage factor $k=2$ instead of 3 and more precise since the uncertainties goes from $6.2\times 10^{-21}$~T$^{-2}$ to $1.2\times 10^{-21}$~T$^{-2}$.

\subsection{Refinement}

\begin{figure}
\centering
\includegraphics[width=\linewidth]{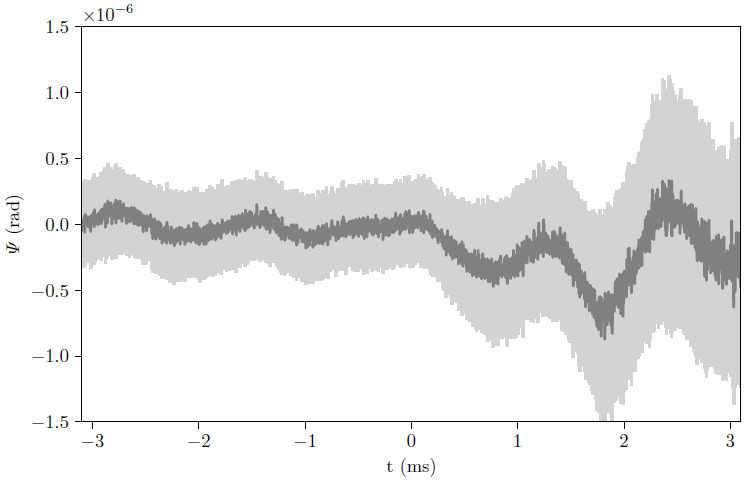}
\caption{Time evolution of the ellipticity and its uncertainties with a coverage factor $k=3$ obtained using the PSD given in \cite{Cadene2014}. We observe an oscillation of frequency 800~Hz.}
\label{fig:Psi800Hz}
\end{figure}

In the previous simulation we use the LSD given by Eq.~\ref{eq:LSDmodel} shown in Fig.~\ref{fig:Bruit}. In this paragraph we will use the full spectrum reported in \cite{Cadene2014} always shown in Fig.~\ref{fig:Bruit}. We simulate data using the same parameters as before, we obtain the result Fig.~\ref{fig:Psi800Hz}. It present a oscillation at 800~Hz that corresponds to the highest peak of Fig.~\ref{fig:Bruit}. If we perform the same analysis as before we obtain a value for the Cotton-Mouton constant of vacuum of $k_{\mathrm{CM}} = (-2.6\pm7.5)\times 10^{-21}$~T$^{-2}$ with a coverage factor $k=3$. This value is compatible with the precedent results. However, the oscillation at 800~Hz is not observed in the experimental data reported in \cite{Cadene2014} (see also Fig.~\ref{fig:don2014} of this paper). Which means that the LSD in Fig.~6 of \cite{Cadene2014} doesn't seem to have been a typical one, corroborating our previous choice to use of LSD to simulate 2014 data.

\section{Conclusion}

In this paper, we present how to generate a data set mimicking true experimental data thanks to a LSD noise spectrum with a Monte Carlo approach. We also introduce a parameter close to least square method one to optimize the central value of the simulation as well as the uncertainties. This approach is very general concerning all optical experiments where a noise spectrum is known. In order to exemplify the method, we apply it to BMV data. We started from the background noise and to match Monte Carlo and real data we added different noise sources: dispersion of birefringence of the Fabry-Perot cavity between experimental magnetic pulses, a random linear increase of the spectral noise caused by acoustic effects and acoustic systematic effect due to the magnetic pulse, which are a confirmation of what we have reported in \cite{Hartman2019}. Thanks to this study we are able to point out the different sources of our experimental noise and thus we can work for their cancellation. Moreover, by simulating some experimental data, we were able to subtract systematic effects and recover the expected effect with a slightly more accurate and precise value than with more traditional analysis methods reported in 2014.  This confirms that the BMV sensitivity of 2014 was essentially limited by acoustic noise. On the other hand our simulation indicates that before the magnetic pulses the values taken by the ellipticity has not to be normally distributed and this cannot be a criterion to exclude data from the analysis.

\section{Appendix : discussion on the intensity ratio formula}
The formula used in 2014 to determine the ellipticity from the intensities is, in our notation
\begin{equation}
P_e(t) = \hat{C}[P_t(t)]\left(\sigma^{2}+(\Gamma+\hat{C}[\Psi(t)])^{2}\right),
\end{equation}
in this appendix we will show that the error performed from the exact expression Eq.~\ref{eq:psi} is negligible. We recall that we have defined the operator $\hat{C}$ associated to the cavity transfer function $H(\nu)$ with a cutoff frequency $\nu_c$ so that $\Psi_{f}(t)=\hat{C}[\Psi(t)]=\mathscr{F}^{-1}[H(\nu)\mathscr{F}[\Psi(t)]]$ where $\mathscr{F}$ is the operator of the Fourier transform.

We consider that the ordinary intensity show a small modulation around a DC value as $P_{t}(t)=P_{t}^{\mathrm{DC}}\left(1+\beta\cos\omega t\right)$ with $\beta \ll 1$. To simplify we will also consider that the magnetic field is oscillating as $B(t)=A\cos\omega_{\mathrm{B}}t$ therefore the ellipticity $\Psi(t)$ can be written as
\begin{align}
    \Psi(t)&=\alpha B^{2}(t)=\frac{\alpha A}{2}\left(1+\cos2\omega_{{B}}t\right)\nonumber\\
    &=\Psi^{{DC}}\left(1+\cos2\omega_{{B}}t\right).
\end{align}
We calculate $P_{e}(t)$ using Eq.~\ref{eq:psi} taking the case $\Psi(t)\ll \Gamma$ so
\begin{equation}
    P_{e}(t)\simeq\sigma^{2}P_{t}(t)+\hat{C}\left(P_{t}(t)[\Gamma^{2}+2\Gamma\Psi(t)]\right),
\end{equation}
injecting our expressions for $P_{t}(t)$ and $\Psi(t)$ we obtain
\begin{align}
    P_{e}(t)={}&\sigma^{2}P_{t}(t)+\hat{C}\left(P_{t}^{{DC}}\left(\Gamma^{2}+2\Gamma\Psi(t)\right)\right)\\
    &+ \hat{C}\left(P_{t}^{{DC}}\beta\cos\omega t\left(\Gamma^{2}+2\Gamma\Psi(t)\right)\right) \nonumber\\
    ={}&\sigma^{2}P_{t}(t)+P_{t}^{{DC}}\Gamma^{2}+2\Gamma P_{t}^{{DC}}\Psi_{f}(t)\nonumber\\
    &+P_{t}^{{DC}}\Gamma^{2}\beta \hat{C}\left(\cos\omega t\right)+2\Gamma P_{t}^{{DC}}\beta \hat{C}\left(\Psi(t)\cos\omega t\right)  \nonumber \\
    ={}&\sigma^{2}P_{t}^{{DC}}+\sigma^{2}P_{t}^{{DC}}\beta\cos\omega t+P_{t}^{{DC}}\Gamma^{2}+2\Gamma P_{t}^{{DC}}\Psi_{f}(t)\nonumber\\
    &+P_{t}^{{DC}}\Gamma^{2}\beta \hat{C}\left(\cos\omega t\right) \nonumber +2\Gamma P_{t}^{{DC}}\beta\Psi^{{DC}}\hat{C}\left(\cos\omega t\right)\nonumber\\
    &+2\Gamma P_{t}^{{DC}}\beta\Psi^{{DC}}\hat{C}\left(\cos\omega t\cos 2\omega_{{B}} t\right). \nonumber
\end{align}
In order to calculate the ellipticity $\Psi^{2014}(t)$ of 2014 we have to determine the inverse of $P_{t,f}(t)=\hat{C}\left(P_{t}(t)\right)$
\begin{equation*}
    \frac{1}{P_{t,f}(t)}=\frac{1}{P_{t}^{{DC}}\left(1+\beta C\left(\cos\omega t\right)\right)}\simeq\frac{1}{P_{t}^{{DC}}}\left(1-\beta C\left(\cos\omega t\right)\right),
\end{equation*}
because $\beta\ll 1$. We calculate the ratio $P_{e}(t)/P_{t,f}(t)$, neglecting of terms in $\beta^{2}$
\begin{align}
    \frac{P_{e}(t)}{P_{t,f}(t)}\simeq {}&\sigma^{2}+\Gamma^{2}+2\Gamma\Psi_{f}(t)+\sigma^{2}\beta\left(\cos\omega t-\hat{C}\left(\cos\omega t\right)\right)\nonumber\\
    &-2\Gamma\beta\Psi_{f}(t)\hat{C}\left(\cos\omega t\right) +2\Gamma\beta\Psi^{\mathrm{DC}}\hat{C}\left(\cos\omega t\right)\nonumber\\
    &+2\Gamma\beta\Psi^{{DC}}\hat{C}\left(\cos\omega t\cos 2\omega_{{B}}t\right)\nonumber \\
    ={}&\sigma^{2}+\Gamma^{2}+2\Gamma\Psi_{f}(t)+\sigma^{2}\beta\left(\cos\omega t-\hat{C}\left(\cos\omega t\right)\right) \nonumber \\
    &+2\Gamma\beta\Psi^{{DC}}\left(\hat{C}\left(\cos\omega t\cos 2\omega_{{B}}t\right)\right.\nonumber\\
    &-\left.\hat{C}\left(\cos\omega t\right)\hat{C}\left(\cos 2\omega_{{B}}\right)\right).
\end{align}
We finaly obtain the ellipticity as it was computed in 2014, that is $\Psi^{2014}(t)=\left(\frac{P_{e}(t)}{P_{t,f}(t)}-\sigma^{2}-\Gamma^{2}\right)/2\Gamma$ and we calcule the error as $\delta(t)=\Psi^{2014}(t)-\Psi_{f}(t)$
\begin{align}
    \delta(t)={}&\frac{\sigma^{2}\beta}{2\Gamma}\left(\cos\omega t-\hat{C}\left(\cos\omega t\right)\right)\\
    &+\beta\Psi^{{DC}}\left(\hat{C}\left(\cos\omega t\cos 2\omega_{{B}}t\right)\right.\nonumber\\
    &-\left.\hat{C}\left(\cos\omega t\right)\hat{C}\left(\cos 2\omega_{{B}}\right)\right)\nonumber\\
    ={}&\frac{\sigma^{2}\beta}{2\Gamma}\left(\cos\omega t-\hat{C}\left(\cos\omega t\right)\right)\\
    &+\beta\Psi^{{DC}}\left(\frac{1}{2}\hat{C}\left(\cos(\omega +2\omega_{{B}})t\right)\right.\nonumber\\
    &+\left.\frac{1}{2}\hat{C}\left(\cos(\omega- 2\omega_{{B}})t\right)-\hat{C}\left(\cos\omega t\right)\hat{C}\left(\cos 2\omega_{{B}}\right)\right).\nonumber
\end{align}
We first study the case where $\omega \ll \omega_{c}$, because a magnetic shot typically last about ten milliseconds we can consider that during the duration of the shot we have $\omega t\ll 1$, so
\begin{equation}
    \delta(t)\simeq\beta\Psi^{{DC}}\left(\hat{C}\left(\cos 2\omega_{{B}}t\right)-\hat{C}\left(\cos 2\omega_{{B}}t\right)\right)=0.
\end{equation}
In the limit $\omega \gg \omega_{c}$, because the cavity filtering on a cosinus is
\begin{equation}
    \hat{C}\left(\cos\omega t\right)=\frac{1}{\sqrt{1+(\omega/\omega_{c})^{2}}}\cos\left(\omega t -\arctan(\omega/\omega_{c})\right),
\end{equation}
so for this limit we have
\begin{equation}
    \delta(t)\simeq\frac{\sigma^{2}\beta}{2\Gamma}\cos\omega t.
\end{equation}
If we use the typical parameters of an experiment like the one in ref.~\cite{Cadene2014}, we found that the amplitude of the error is of the order of $10^{-9}$.

We can therefore consider that the formula used in 2014, although incorrect, presents a negligible error.

\section{Acknowledgments}
This research has been partially supported by ANR (Grant No. ANR-14-CE32-0006).  We thank all the members of the BMV collaboration. We specially thank M. Hartman and A. Charry and all the authors of \cite{Cadene2014}.

\section{Authors contributions}
All the authors were involved in the preparation of the manuscript.
All the authors have read and approved the final manuscript.

\bibliographystyle{unsrt}  


\end{document}